\documentclass{iopconfser}
\usepackage{amsmath, amssymb, hyperref}
\usepackage{orcidlink}
\begin{document}

\title{Search for growing angular modes in ultracompact boson star evolutions}

\author{Seppe J. Staelens \orcidlink{0000-0002-1262-1600} $^{1,2}$}

\affil{$^1$DAMTP, Centre for Mathematical Sciences,
University of Cambridge, Wilberforce Road, Cambridge CB3 0WA, UK\\ $^2$Leuven Gravity Institute, KU Leuven,
Celestijnenlaan 200D box 2415, 3001 Leuven, Belgium}

\email{ss3033@cam.ac.uk}

\begin{abstract}
Recent fully non-linear simulations of ultracompact spherically symmetric boson stars have presented evidence for their long-term stability.  All observed dynamics could be mainly attributed to the fundamental radial mode. In this work, we additionally decompose some of the data in spherical harmonics, providing a first step towards the characterization of non-spherical modes present.
\end{abstract}

\section{Introduction}

Over the past decade, gravitational wave (GW) detections \cite{Abbott:2016blz, KAGRA:2021vkt} and millimeter observations of black hole (BH) environments \cite{EventHorizonTelescope:2019dse, EventHorizonTelescope:2022wkp} have opened the door to strong tests of the Kerr-paradigm: are astrophysical BHs (on all mass scales) described by the Kerr metric in General Relativity (GR)? Theoretical considerations
have led physicists to explore alternative avenues. On the one hand, the quest for a high-energy unification of gravity and quantum physics has given rise to a plethora of beyond-GR theories and their associated BH solutions. At the same time, the community has explored alternatives to BH solutions, the so-called \textit{ exotic compact objects} (ECOs), in order to avoid potential problems associated with the event horizon \cite{Mathur:2009ip} and/or the singularity \cite{Hawking:1973uf}.

For any of these ECOs to be considered serious contenders, they should surpass several benchmarks. First of all, their observational predictions should fall within the current constraints; a field of research which has seen a lot of recent work - see e.g. \cite{bambi2025blackholemimickerstheory} for a summary.
Due to this prerequisite, special interest has been devoted to objects that possess one or more light rings (LRs), bound null geodesics that strongly influence the optical appearance of accreting BHs \cite{bardeen1973timelike, lupsasca_beginners_2024} and are intimately connected with quasinormal modes \cite{Cardoso:2008bp, Jusufi:2019ltj, Pedrotti:2025idg}. Throughout the rest of this work, we refer to such objects with LRs as \textit{ultracompact}. Additionally, we refer to ultracompact ECOs as BH mimickers.

Second, any such object should be dynamically viable, meaning that there should be a plausible formation mechanism, and the object should be stable against the ubiquitous perturbations astrophysical objects experience. It is with respect to the latter that the a priori promising ultracompact ECOs run into complications. 
ECOs that could form through some process of gravitational collapse acquire LRs in pairs once they become ultracompact, both an unstable and a stable one \cite{Cunha:2017qtt}. Waves propagating in the spacetime of such a BH mimicker exhibit logarithmic decay in the potential well around the stable LR at linear level \cite{Keir:2014oka}. A similar result was obtained for anti-de Sitter, which turns out to source an instability \cite{Bizon:2011gg, Holzegel:2011uu, Holzegel:2013kna}. Therefore, it was deemed plausible that such ultracompact ECOs may also be nonlinearly unstable, ruling them out as effective BH mimickers \cite{Cardoso:2014sna}.

As opposed to the case of anti-de Sitter, this putative instability has not been proven, although different analyses have been put forward: analytical \cite{Redondo-Yuste:2025hlv, Cunha:2025oeu} and numerical \cite{Cunha:2022gde, Benomio:2024lev, Guo:2024cts, Evstafyeva:2025assessing} results provide arguments for and against the instability, thus leaving the question open as of today. In particular, boson stars (BSs, see \cite{liebling2023dynamical} for a review) provide an excellent laboratory to investigate the proposed LR instability in fully non-linear numerical evolutions. Initial results suggested that the instability is present and effective in destroying the LR structure in rotating BSs \cite{Cunha:2022gde}, though recent results have challenged their conclusion \cite{Evstafyeva:2025assessing}. The case of ultracompact solitonic BSs that are spherically symmetric has recently been investigated in Ref.~\cite{Marks:2025longterm}, where two explicit models have been evolved over a (numerically) long timescale, solving the full non-linear Einstein equations with and without symmetry assumptions. The study finds no evidence of an instability on the timescales considered, and shows that the LR structure is preserved throughout the evolution. The observed dynamics can all be explained by radial perturbation theory, which also suggests stability for these particular models. 

The present work builds on the latter, and aims to provide further evidence for the absence of any instability in the simulations. The single-BS spacetimes are evolved in time in full $3D$ using {\sc ExoZvezda}~\cite{exozvezda}, a code for exotic matter built on {\sc GRChombo}~\cite{Radia:2021smk,Andrade:2021rbd}, which evolves $3D$ domains with adaptive mesh refinement provided by {\sc chombo} \cite{chombo}. We use the conformal covariant $Z4$ (CCZ4) \cite{Alic:2011gg} formulation, and the setup of the simulations is identical to the ones in Ref.~\cite{Marks:2025longterm}, unless stated otherwise. We denote their BS models as \texttt{S06A044} (\texttt{S08A06}), i.e. having $\sigma = 0.06$ (0.04), $A_0 = 0.044$ (0.06) and scalar mass $\mu$. We use units such that $G = c = \hbar = 1$.

\section{Angular decomposition}

{\textbf{\textit{Method}}---} We extract the modulus of the scalar field $\varphi$ and the Adiabatic Effective Potential (AEP)\footnote{The effective potential, as defined for stationary and spherically / axially symmetric spacetimes, can be well approximated by the AEP when deviations from these symmetries are small.} \cite{Cunha:2022gde, Marks:2025longterm}, given by $H_{\text{eff}}(r) = \sqrt{\langle \alpha^2 \rangle - \langle \gamma_{ij} \beta^i \beta^j\rangle} / r$, on $N_\theta \times N_\phi$ points ($\theta$ and $\phi$ are the polar and azimuthal angle respectively) on $N_R$ spheres of coordinate radius $R$, at fixed simulation time intervals $\Delta T$. We decompose these extractions into spherical harmonics, using \texttt{Python} and \texttt{scipy} \cite{2020SciPy-NMeth}, and consider each mode as a function of time, normalized by the amplitude of the (0,0)-mode. To assess whether any modes are significantly growing in time, we perform a linear fit on the base-10 logarithm of the mode amplitude in time: this should detect modes that are exponentially growing\footnote{Note that exponential growth is typically associated with a linear instability, while the putative LR instability is expected to be non-linear. Given the uncertainty of how the instability would manifest itself, looking for exponentially growing modes means expecting a 'worst-case' scenario.} with respect to the (0,0)-mode. We only keep modes for which the best-fit parameter for the slope is positive, and perform a $t$-test\footnote{Note that this assumes that the variance of all data points is the same - an assumption that may not be true here, but good enough to assess which modes are most significantly growing.} using \texttt{statsmodels} \cite{seabold2010statsmodels} that rejects the null hypothesis that the mode is not growing in time (i.e. the slope is zero) if the $p$-value is smaller than 5\%. 
Extraction of the scalar field amplitude was added later and does not always cover the entire simulation
We expect physically growing modes to affect both the scalar field and effective potential, and therefore look for consistency. We typically neglect the start of the simulation $\mu t \leq 500$, where initial dynamics are expected to grow all modes. Given the octant symmetry employed in the simulations, the analysis is not sensitive to modes with odd $l,m$. We have verified that the recovered mode amplitude does not change significantly with varying angular resolution $N_\theta, N_\phi$.

\noindent{\textbf{\textit{Baseline}}---} The spherical harmonics $Y_{lm} (\theta,\phi)$ satisfy orthogonality relations on the unit sphere $S^2$: $ \int_{S^2} Y_{lm}(\theta,\phi) Y_{l'm'}(\theta,\phi) \:\text{d}\Omega = \delta_{ll'} \delta_{mm'}$.
As we are extracting a finite number of points on the sphere, however, the integral has to be approximated by a sum, and the orthogonality relations do not hold exactly. As a consequence, some of the modes extracted in our analysis could be artificially large due to the numerical integration. Therefore, we first explore what modes our analysis predicts for data that are perfectly spherically symmetric, i.e. a $(0,0)$-mode. The extracted modes with $l$ even and $m=0$ are significant compared to the (0,0)-mode, on the order $\sim 10^{-3}$.
They are recovered at the same amplitude in the analysis of the actual simulation data, and are therefore deemed uninformative. All other modes have an amplitude 14-19 orders of magnitude smaller than that of the $(0,0)$-mode, and therefore are not strongly affected by the numerical approximation. Furthermore, we also only consider extracted modes with $l < N_\theta, m <N_\phi$, as higher modes are not resolved by the discretisation. A floor value of $10^{-12}$ is used below which (relative) mode values are not extracted.
\begin{figure}[t]
    \centering
    \includegraphics[width=\linewidth]{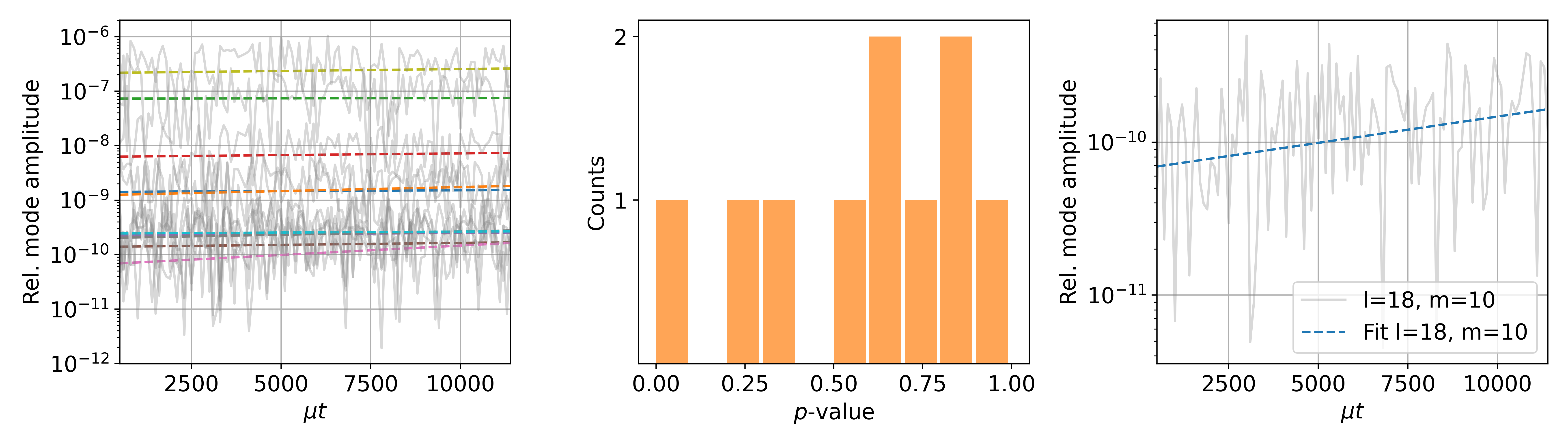}
    \caption{Analysis for the effective potential $H_{\text{eff}}$ in \texttt{S06A044} at resolution $\mu h=1/8$, using $(N_\theta, N_\phi) = (32,24)$ at $R=17.5$. \textit{(left)} Extracted modes and linear fits, for those with a positive slope, \textit{(middle)} histogram of the associated $p$-values and \textit{(right)} the $(18,10)$-mode that is significantly increasing.}
    \label{fig: s06_1over8}
\end{figure}

\section{Results}

\noindent{\texttt{S06A044}---} Figure~\ref{fig: s06_1over8} shows the results of the analysis of the effective potential for the evolution at resolution $\mu h=1/8$, using $(N_\theta, N_\phi) = (32,24)$ at $R=17.5$ (approximately corresponding to areal radius $r=28.9$, around which the stable LR location oscillates \cite{Marks:2025longterm}). The histogram of $p$-values shows that one mode, $(18,10)$, has a slope that is significantly positive - shown in the rightmost panel.
An analysis of the scalar field amplitude (at times $\mu t\gtrsim 8700$ only) does not recover this mode, but rather identifies two different modes - $(18,18)$ and $(12,2)$ -, neither of which are flagged by $H_{\text{eff}}$. Upon restricting the latter to the same time interval, the $(18,18)$-mode is almost significant, with a $p$-value of about 5.2\%. The $(18,10)$-mode is not flagged by the scalar field analysis.

We have done the same analysis on the effective potential in the simulation with resolution $\mu h = 1/12$. Due to the computational cost, extractions were only obtained at times $10,000 \lesssim \mu t \lesssim 12,200$ (continuing from a checkpoint of Ref.~\cite{Marks:2025longterm}). If we extract the modes at a time interval $\mu \Delta T = 100$, only the mode $(14,10)$ is flagged as significantly growing; upon changing to $\mu \Delta T = 50$, the analysis now suggests $(16,4)$.

Overall, we note that the different analyses do not provide consistent results on which modes are significantly growing. In addition to the overall low relative amplitudes ($\lesssim 10^{-6}$) of the modes and the visible scatter of the data, the results suggest that the recovered growing modes depend on the parameters of the simulation and analysis. We can therefore not confidently say that these modes are physically growing, and are likely attributed to numerics.

\begin{figure}[t]
    \centering
    \includegraphics[width=\linewidth]{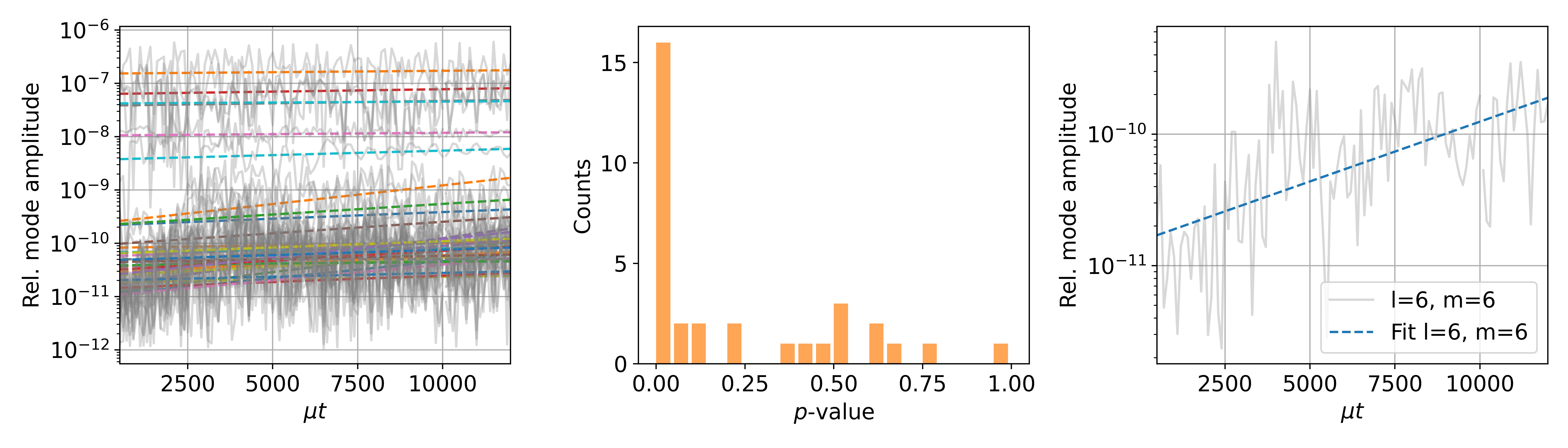}
    \caption{Analysis for the scalar field amplitude $|\varphi|$ in \texttt{S08A06} at resolution $\mu h=1/12$, using $(N_\theta, N_\phi) = (32,24)$ at $R=9.8$. \textit{(left)} Extracted modes and linear fits, for those with a positive slope, \textit{(middle)} histogram of the associated $p$-values and \textit{(right)} the $(6,6)$-mode that is most significantly increasing.}
    \label{fig: s08_1over12_Phi}
\end{figure}

\noindent{\texttt{S08A06}---} We analyze a simulation with resolution $\mu h=1/12$, using $(N_\theta, N_\phi) = (32,24)$ at $R=9.8$ ($\mu r\approx16$), over a time range $\mu t \in [500, 12000]$ at intervals $\mu \Delta T = 100$.
While $H_{\text{eff}}$ suggests 4 growing modes (two marginally), the scalar field amplitude $|\varphi|$ suggests 16: 3 of these - (2,2), (22,22) and (18,6) - are flagged by both. In $|\varphi|$, all growing modes have amplitudes $\lesssim 10^{-9}$, while the larger modes are all estimated as not growing. The left and right panels in Fig.~\ref{fig: s08_1over12_Phi} suggest that, while an initial growth is apparent for these small modes, a saturation is reached somewhere in $\mu t \in [3000, 4000]$. Increasing the starting time of the analysis to $\mu t = 4000$ supports this conclusion: 5 (2) modes are marginally flagged in $|\varphi|$ ($H_{\text{eff}})$.
We conclude that, while a strong initial growth is detected for the smallest modes in the scalar field amplitude, the data at later times suggest saturation. Thus, we do not find strong evidence of modes that are growing throughout the entire simulation.

\section{Conclusion}
Although individual analyses indicate that a few modes may be growing over (small portions of) time, no consistent growth is found across the board: changing the time range, the frequency of extraction, the simulation resolution and the extraction radius can result in different modes being flagged. Additionally, upon extracting modes of both the effective potential and the scalar field amplitude, the results do not exhibit coherent growth in both: few exceptions are found - $(18,18)$ in the \texttt{S06A044}-run at resolution $\mu h = 1/8$ and $(22,12)$ in the \texttt{S08A06}-run at $\mu h = 1/8$ - where both analyses hint at growth, and even then only when restricted to specific time ranges.

If a physical instability were present, we would expect it to trigger growth of several modes (likely with high $l,m$), rather than a single specific one. Additionally, we would expect consistency across different time ranges, simulation resolutions (up to convergence) and diagnostics (in this case, the effective potential and scalar field amplitude). We conclude that this is not the picture that our analysis paints.

It should be noted that our analysis can be improved in several ways, and may not have the precision required to resolve potentially growing modes that remain small on the timescale of the simulation. Additionally, we could only probe up to finite values of $l$ and $m$. Overall, the results here are in agreement with the conclusions of Ref.~\cite{Marks:2025longterm} - in particular Figs.~7 and 8 of the Supplemental Material - that no sign of a physical instability is recovered in our simulations.

\section*{Acknowledgements}
The author acknowledges useful discussions with G. A. Marks and U. Sperhake. S. J. S. is supported by the Centre for Doctoral Training at the University of Cambridge funded through STFC. We acknowledge support by the NSF Grant Nos. PHY-090003, PHY-1626190 and PHY-2110594, DiRAC projects ACTP284 and ACTP238, STFC capital Grants Nos. ST/P002307/1, ST/R002452/1, ST/I006285/1 and ST/V005618/1, STFC operations Grant No. ST/R00689X/1. Computations were done on the CSD3, Fawcett (Cambridge) and Cosma (Durham) clusters.

\vspace{7mm}
\bibliography{ref}

\end{document}